%% file: polyhedralGFT.tex
\documentclass{PoS}

\usepackage{amsthm, amsmath, amssymb, mathrsfs}
\usepackage{pgf,tikz}
\usepackage{cite}

\include{macros}

\title{Group field theories generating polyhedral complexes}

\ShortTitle{GFTs generating polyhedral complexes}

\author{\speaker{Johannes Th\"urigen}\thanks{based on work in collaboration with Daniele Oriti and James Ryan \cite{\ORT}}\\
         Max-Planck Institute for Gravitational Physics (Albert-Einstein Institute), Potsdam, Germany\\
         E-mail: \email{Johannes.Thuerigen@aei.mpg.de}}

\abstract{Group field theories are a generalization of matrix models which provide both a second quantized reformulation of loop quantum gravity as well as generating functions for spin foam models. 
While states in canonical loop quantum gravity, in the traditional continuum setting, are based on graphs with vertices of arbitrary valence, group field theories have been defined so far in a simplicial setting such that states have support only on graphs of fixed valency. 
This has led to the question whether group field theory can indeed cover the whole state space of loop quantum gravity.
In this contribution based on \cite{\ORT} I present two new classes of group field theories which satisfy this objective: i) a straightforward, but rather formal generalization to multiple fields, one for each valency and ii) a simplicial group field theory which effectively covers the larger state space through a dual weighting, a technique common in matrix and tensor models. 
To this end I will further discuss in some detail the combinatorial structure of the complexes generated by the group field theory partition function.
The new group field theories do not only strengthen the links between the mentioned quantum gravity approaches but, broadening the theory space of group field theories, they might also prove useful in the investigation of renormalizability.
}

\FullConference{Frontiers of Fundamental Physics 14 - FFP14,\\
                15-18 July 2014\\
                Aix Marseille University (AMU) Saint-Charles Campus, Marseille }
		
\begin{document}


\section*{Motivation}

Loop quantum gravity (LQG) \cite{\lqgT,Rovelli:2011tk} , spin foam (SF) models \cite{Perez:2013uz} and group field theory (GFT) \cite{Oriti:2012wt,Krajewski:2012wm,Oriti:2014wf} are three closely related approaches to a quantum theory of gravity.
They share that the degrees of freedom are algebraic data based on certain combinatorial structures and SF models and GFT can be understood as providing dynamics for LQG.
A major difference between the approaches consists in the role and structure of these combinatorics.
The quantum states of traditional LQG are arbitrary closed graphs embedded in a smooth $(\std-1)$-dimensional spatial 
manifold and related by cylindrical consistency.
Spin foam models are originally based on triangulations of a given topological $\std$-manifold such that boundary states are based on $\std$-regular graphs, \ie all vertices are $\std$-valent.
But SF models can also be defined on other cellular decompositions and the amplitudes usually depend only on their 2-skeleton which has lead to extensions to more general 2-complexes with arbitrary boundary graphs \cite{\KKL}.
GFTs can be understood as a completion of complex-dependent SF models, 
providing a sum of SF amplitudes over a class of combinatorial complexes of various topologies, generated from $\std$-simplicial building blocks. Boundary states are thus based on $\std$-regular graphs.

Here I would like to address this difference in the combinatorial structure and show that there are also classes of GFTs combinatorially fully compatible with LQG and SF models.
The challenge is to find GFTs which generate complexes that allow for boundary graphs of arbitrary valencies.

Explicating in detail the combinatorial structure of standard GFT in section 1 will lay the ground to present two strategies for such an extension: 
i) In section 2, I will extend the field space to a set of fields with various numbers of group arguments, thus creating boundary vertices of various valency. Though rather formal, such multi-field GFTs are the direct counterpart to the KKL extension of spin foams \cite{\KKL}.
ii) In section 3, I will therefore present a more efficient generalization in terms of a dual-weighting mechanism. 
An extra label on a standard simplicial GFT field allows to distinguish real from virtual cells and the dual weighting implements dynamically a contraction of the virtual structures resulting in effective amplitudes which are exactly the same as in the multi-field GFT.


\section{The combinatorial structure of group field theory}

The common notion of GFT is that of a quantum field theory on group manifolds with a particular kind of  non-local interaction vertices \cite{Oriti:2012wt,Krajewski:2012wm,Oriti:2014wf}.
More precisely, a group field is a function $\phi:G^{\times\copies}\ra\R$ of $\copies\in\N$ copies of a Lie group $G$ and the GFT is defined by a partition function
\begin{equation}
 \label{eq:partition}
Z_{\gft} = \int \Dcal\phi\; e^{-S[\phi]}\;,
\end{equation}
where $\Dcal\phi$ denotes a (formal) measure on the space of group fields and the action is of the form
\begin{equation}
   S[\phi] = \frac12\int [\d g]\; \phi(g_1)\;\Kbb(g_1, g_2)\;\phi(g_2) + \sum_{i\in I}\lambda_i \int [\d g]\;\Vbb_i\big(\{g_j\}_{J_i}\big)\;\prod_{j\in J_i}\phi(g_{j})\;.
 \label{action}
\end{equation}
Therein, the kinetic kernel 
$ \Kbb(g_1, g_2)  = \Kbb(g_{11},g_{21};\dots;g_{1\copies},g_{2\copies}) $ 
is a function on $G^{2\copies}$ pairing the arguments;
the vertex (interaction) kernels $\Vbb_i$ with couplings $\lambda_i$
are functions on $G^{\copies|J_i|}$ for finite sets $J_i$ and have a combinatorially non-local structure.
This means that the kernels $\Vbb_i$ do not impose coincidence of points in the group space $G^{\times\copies}$ but that the totality of the $\copies \cdot |J_i|$ field arguments is partitioned into pairs convoluted by the kernel, 
\[\label{convolution}
 \Vbb\big(\{g_j\}_J\big)  = \Vbb\big( \{g_{ja}g_{kb}^{-1}\}\big)
\]
where $j,k\in J$ and $a,b\in\{1,...,\copies\}$.
Observables $O[\phi]$ commonly have kernels of the same type.

For the evaluation of expectation values of such quantum observables $O[\phi]$, a perturbative expansion with respect to the coupling constants $\{\lambda_i\}_I$ leads to a series of Gaussian integrals evaluated through Wick contraction which are catalogued by Feynman diagrams $\fdiagram$,
\begin{eqnarray}\label{expansion}
  \qbra O\qket_{\gft} &=& 
  \frac{1}{Z_{\gft}} \int \Dcal\phi\;O[\phi] \sum_{\{c_i\}_I} \prod_{i\in I}\frac{1}{c_i!}\Bigg[ \lambda_i \int [\d g]\;\Vbb_i\Big(\{g_j\}_{J_i}\Big)\;\prod_{j\in J_i}\phi(g_{j})\Bigg]^{c_i} e^{-\frac12\int[\d g]\;\phi(g_1)\;\Kbb(g_1,g_2)\;\phi(g_2)}\nonumber\\
&=& \sum_{\fdiagram} \frac{1}{\sym(\fdiagram)} A(\fdiagram;\{\lambda_i\}_I)\;,
\end{eqnarray}
where $\sym(\fdiagram)$ are the combinatorial factors related to the automorphism group of the Feynman diagram $\fdiagram$  and $A(\fdiagram;\{\lambda_i\}_I)$ is the weight of $\fdiagram$ in the series. These weights are constructed by convolving (in group space) propagators $\Pbb = \Kbb^{-1}$ and interaction kernels $\Vbb_i$.

\begin{figure}
  \centering
  \includegraphics{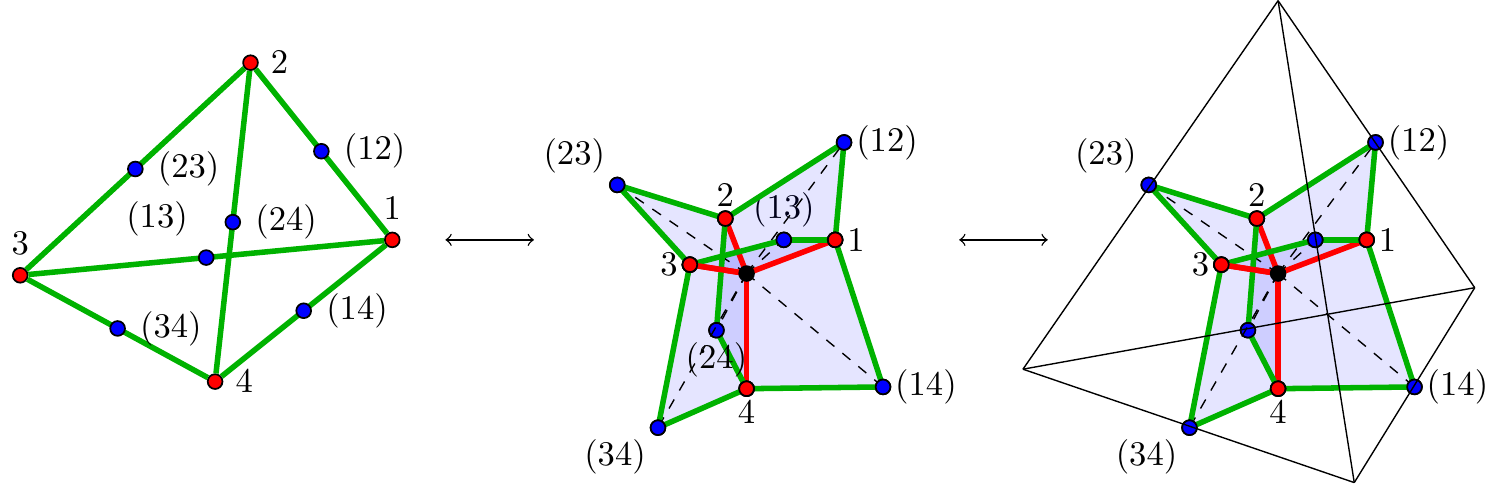}
\caption{\label{fig:dualtet} A bisected $\copies$-regular graph, its corresponding spin foam atom, and a $\copies$-dimensional extension which, in this example, is the dual of the tetrahedron.}
\end{figure}

The essential point for further clarification of the structure of these diagrams is to notice that the specific non-locality of each vertex is captured by a boundary graph. 
In the interaction term in \eqref{action}, each group field term $\phi(g_j) = \phi(g_{j1},\dots,g_{j\copies})$ can be represented by a graph consisting of a $\copies$-valent vertex $\vb_j$ connected to $\copies$ univalent vertices $\vh_{j1},\dots,\vh_{j\copies}$, coined a \emph{patch} and denoted $\bp_{\vb_j}$.
A bisected graph $\bbg$ represents then the convolutions in \eqref{convolution} by identifying for each argument $g_{ja}g_{kb}^{-1}$ the univalent vertices $\vh_{ja},\vh_{kb}$ into a bivalent vertex $\vh_{ij}$. 
One may further understand the graph as the boundary $\bbg = \bs\sfa$ of a two-dimensional complex $\sfa$ with a single internal vertex $v$ in a unique way, \ie connecting $v$ to all vertices in $\bbg$ and adding a face $(v\vb\vh)$ for every edge $(\vb\vh)$ in $\bbg$ (see \fig{dualtet} for illustration).
Such a one-vertex two-complex $\sfa$ is called a \emph{(spin foam) atom}.

\begin{figure}
  \centering
  \includegraphics{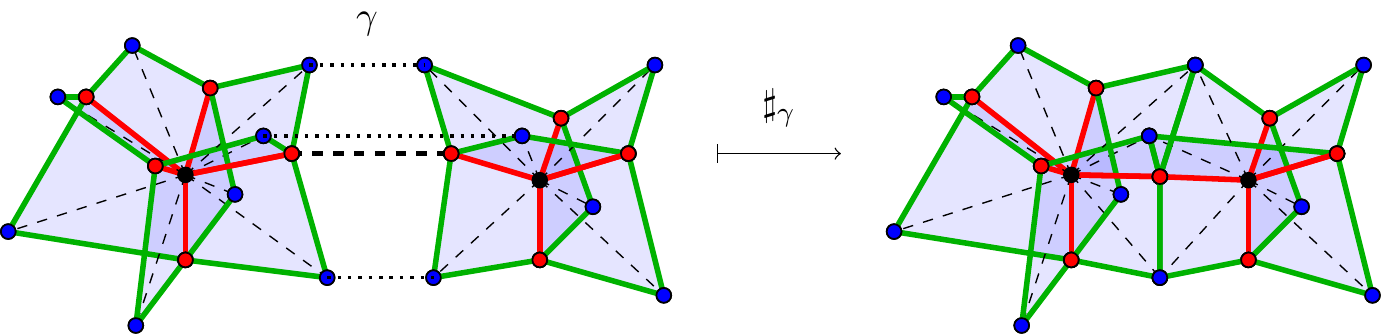}
\caption{\label{fig:molecule} The bonding $\sharp_{\gm}$ of two atoms along an identification of patches $\gm$.}
\end{figure}

In this way, the GFT Feynman diagrams in the perturbative sum \eqref{expansion} 
have the structure of two-complexes because Wick contractions effect bondings of such atoms along patches.
In the graph picture, the pairing of group arguments in the kinetic kernel $\Kbb$ induces a bijection $\gm:\bp_{\vb_1}\ra\bp_{\vb_2}$ between the patches representing the two group field terms $\phi(g_1)$ and $\phi(g_2)$.
The combinatorial structure of a term in the perturbative sum \eqref{expansion} is then a collection of spin foam atoms, one for each vertex kernel, quotiented by a set of bonding maps, one for each Wick contraction (\fig{molecule}).
Because of this construction such a two-complex will be called a \emph{(spin foam) molecule}.

Standard GFTs, as described by the action \eqref{action}, generate a peculiar kind of complexes determined by the regular-graph structure of the interaction vertices.
In the set of closed bisected multigraphs $\bbgs$, only graphs in the subset $\bbgs_\rl$ of $\copies$-regular, loopless graphs are possible for the interaction kernels in \eqref{action} since these have to be constructed from the single patch corresponding to the group field with $\copies$ arguments.
Accordingly, the class of molecules which can be obtained from atoms based on $\bbgs_\rl$ is denoted $\sfrs_\rl$.

A special case is \emph{simplicial} GFT defined by the single interaction vertex based on the combinatorics of the $\copies$-simplex.
The corresponding vertex graph $\bbg_\rs$ is the bisected version of the complete graph with $\copies + 1$ vertices (\fig{dualtet}) and the resulting molecules are a subset $\sfrs_\rs \In \sfrs_\rl$.
Since every atom in such a simplicial molecule has a canonical extension to the $\copies$-simplex, it is natural to set $\copies = \std$ and understand $\sfrs_\drs$ as $\std$-dimensional discrete spacetimes.
But note that in general simplicial molecules are not abstract simplicial complexes due to various kinds of loops \cite{Gurau:2010iu,thesis}.
Well-behaved simplicial complexes preventing such degeneracies can be obtained in a modified version of simplicial GFT, distinguished by a colouring \cite{Gurau:2010iu,Gurau:2012hl}.
Integrating out all but one colour one even arrives at a GFT of the general action type \eqref{action} whose vertex atoms have all an extension to $\std$-polytopes due to their effective construction as triangulations \cite{Bonzom:2012bg}.

Even though the GFTs described provide already a broad class of theories, in any case the combinatorial boundary structure is still simplicial.
In particular, only observables and states based on $\copies$-regular loopless boundary graphs $\bbgs_\rl\In\bbgs$ are possible since they are defined in terms of the field $\phi$ with $\copies$ group arguments.
To match the larger space of states of LQG based on the whole of graphs $\bbgs$, new classes of theories have to be introduced.
This is the topic of the next two sections.


\section{Multi-field group field theory}

The most straightforward way to generalize GFT to include observables and states on arbitrary graphs is to extend the field space \cite{Reisenberger:2001hd}.
Though from a QFT point of view a rather unattractive strategy, I present it here to point out the possibility of such a theory and discuss its combinatorics.

Since a patch corresponds to a group field in the GFT, one has to extend the field space to a group field for each kind of patch needed to construct all boundary graphs $\bbgs$.
Let the set of all such patches be $\bps$. 
A multi-field (MF) group field theory is then determined by a set of group fields $\Phi_\sub = \{\phi_\bp\}_{\bps_\sub}$, $\bps_\sub \In \bps$,  which are functions of group elements $g_{\vb\vh}$, one for each edge $(\vb\vh)\in\bp$. 
Denoting $\bbgs_\sub \In \bbgs$ the set of all graphs constructed from $\bps_\sub$ and $\Vb$ the set of all but the bisecting vertices in a boundary graph $\bbg\in\bbgs$, the action is then
\begin{equation}
  \label{eq:multiaction}
  S[\Phi_{\sub}] 
  =  \sum_{\bp\in \bps_{\sub}} \frac12 \int [\d g]\; \phi_{\bp}(g_{\vb_1})\; \Kbb_{\bp}(g_{\vb_1},g_{\vb_2})\;\phi_{\bp}(g_{\vb_2}) 
  + \sum_{\bbg\in\bbgs_{\sub}}  \lambda_{\bbg} \int [\d g]\;\Vbb_{\bbg}(\{g_{\vb}\}_{\Vbar})\prod_{\vb\in\Vbar}\phi_{\bp}(g_{\vb})\;.
\end{equation}
Like in corresponding spin foam models \cite{\KKL}, there is no direct spacetime interpretation for the molecules generated in the perturbative sum for this action.
One might choose a subset of $\bbgs_\sub$ for interactions and observables in which each atom can be extended to the dual of a $\std$-polytope.
In this way one obtains a $\std$-complex for each molecule in the expansion, though in analogy to the simplicial case these complexes are abstract polyhedral complexes only in a generalized sense \cite{thesis}.

To cover the whole LQG state space, the infinite number of fields $\Phi_\sub = \{\phi_\bp\}_\bps$ is necessary. Such a theory will likely remain on a formal level.
In particular, infinitely many interactions are needed for non-trivial dynamics for each field.
Still, this is the theory generating all possible spin foam molecules $\sfrs$, 
and, with appropriate kinetic kernels $\Kbb_\bp$, it generates the amplitudes of the KKL extension \cite{\KKL} of the EPRL \cite{\EPRL} and similar spin foam models \cite{\FK,\BO} (see \cite{\ORT} for details).

For a construction based on those GFTs which are currently analytically tractable it is necessary to rearrange the molecules discussed in this section in terms of the simplicial molecules of the preceding section. 
In the following section I show how this can be done and implemented dynamically using a dual weighting in the GFT action.


\section{Dually-weighted group field theory}

\begin{figure}[htb]
    \centering
    \includegraphics{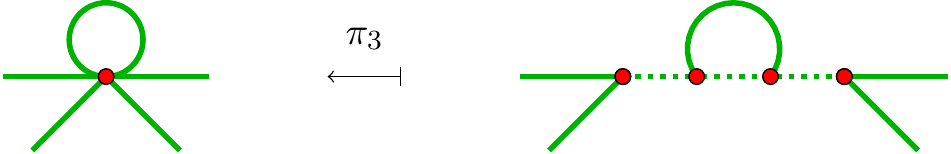}
    \caption{\label{fig:3val} A vertex in an arbitrary graph obtained from a labelled 3-valent graph via contraction.}
  \end{figure}

The crucial idea to create arbitrary boundary graphs in a more efficient way is to distinguish between virtual and real edges and obtain arbitrary graphs from regular ones by contraction of the virtual edges.
Boundary graphs $\bbgst$ with such an edge labelling are obtained from labelled patches $\bpst$ by restricting the allowed identifications of univalent vertices $\vh_{ja}, \vh_{kb}$ to the case where both are adjacent to an edge with the same label.  
Then one can easily check \cite{\ORT} that any graph $\bbg\in\bbgs$ is in the image of the projection $\pi_\rl : \bbgst_\rl \ra \bbgs$ defined as contraction of all virtual edges in a labelled $\copies$-regular, loopless graph for $\copies$ odd (while for $\copies$ even the image consists of all graphs with vertices of even valency). An example is illustrated in \fig{3val}.

\begin{figure}
    \centering
        \includegraphics{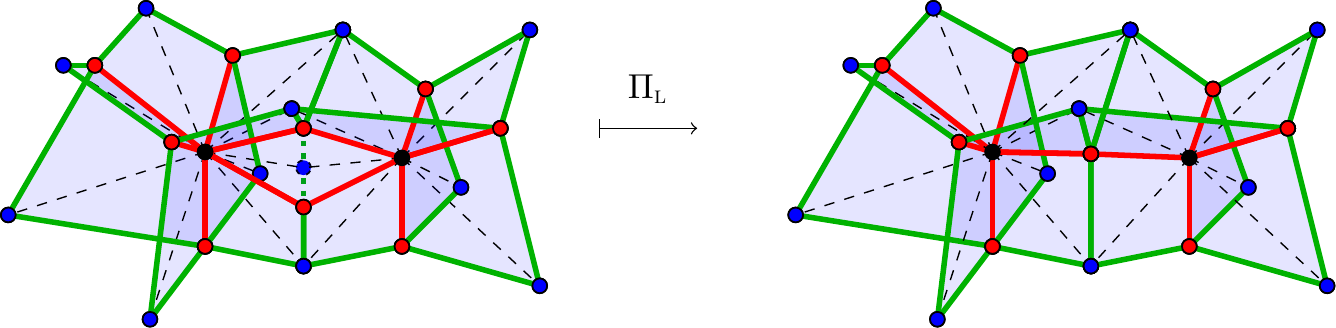}
    \caption{\label{fig:molecule-contraction} Contraction move with respect to a vertex $\vh$ incident to two virtual edges in a molecule.}
 \end{figure}

In terms of these contractions, any spin foam molecule can be obtained from a molecule constructed from labelled regular graphs.
The contraction map $\pi_\rl$ has a natural extension to a map $\Pi_\rl$ on the labelled molecules $\sfrst_\rl$ resulting from bonding atoms induced by $\bbgst_\rl$ (\fig{molecule-contraction}).
To avoid branching one further restricts to $\sfrst_{\copies,\lnb} \In \sfrst_\rl$, the molecules where bisecting vertices $\vh$ are bivalent also after bonding.
From the surjectivity of $\pi_\rl$ follows then that $\Pi_\rl(\sfrst_{\copies,\lnb})=\sfrs$.
One can show further that even a restriction to labelled simplicial molecules $\sfrt\in\sfrst_{\copies,\snb}$ is enough to obtain a molecule $\sfr=\Pi_\rs(\sfrt) \in\sfrs$ with boundary  $\bs\sfr = \bbg$ for any graph $\bbg\in\bbgs$ \cite{\ORT}.

Based on these combinatorics facts it is possible to define a labelled simplicial GFT which effectively covers  observables and states on arbitrary graphs, using a dual-weighting mechanism to implement the contractions dynamically.
To this end, the $\copies$ arguments of the field $\phi(g_\vb;m_\vb)$ are a combination of a group element $g_{\vb\vh}$ and an integer $m_{\vb\vh}\in\{0,1,\dots,M\}$.
If $m_{\vb\vh}=0$ the corresponding edge $(\vb\vh)$ is real, otherwise virtual.
A kinetic kernel with a family of dual-weighting matrices $\{A_M\}$ 
\[
\Kbb(g_1,g_2;m_1,m_2) = \Kbbb(g_1,g_2;m_1,m_2)\;\prod_{j=1}^\copies 
\left(
    \begin{array}{c|c}
      1&0\\ \hline
      0& A_M^{-1}
    \end{array}
    \right)_{m_{1j} m_{2j}}
    \textrm{where}\quad 
     \lim_{M\rightarrow\infty}
  \tr\left( (A_M)^k \right) = \delta_{k,2}
\]
guarantees then that in the large-$M$ limit the perturbative expansion \eqref{expansion} of the theory is a sum only over molecules $\sfrst_{\copies,\snb}$. 
With an appropriate kernel $\Kbbb$ which distinguishes exactly whether an edge is real or virtual, the effective large-$M$ amplitudes only depend on the contracted molecules $\Pi_\rs(\sfrst_{\copies,\snb})$ and match exactly the ones of the multi-field GFT \cite{\ORT}, and thus the amplitudes of the KKL extension \cite{\KKL} of spin foam models \cite{\EPRL,\FK,\BO}.


\section*{Conclusions}
In this contribution I have addressed the goal of a generalization of GFT to be compatible with LQG in three steps.
First, I have clarified the combinatorial structure underlying the amplitudes of perturbative GFT using the notion of spin foam atoms and molecules and discussed their possible spacetime interpretation.
Then I have laid out the details of the straightforward generalization to a multi-field theory which can cover on a formal level arbitrary such molecules.
Finally I have shown how to obtain the same dynamics using a dual-weighting mechanism on a simplicial GFT.

Improving the relations between GFT, SF models and LQG and broadening the GFT theory space, these results introduce two obvious research questions.
i) While all gravitational models are based on a simplicial version of simplicity constraints and the resulting edge amplitudes can be implemented in the presented GFTs either on the molecules $\sfrs$, or maybe more meaningful on the simplicial molecules $\sfrst_{\copies,\snb}$, one would expect a genuine polyhedral version of the constraints to be more appropriate on these molecules, which has not been addressed so far.
ii) From the GFT perspective, it will be most interesting to investigate the field-theoretic properties of the new classes of theories, in particular their large-$N$ limit \cite{Gurau:2012hl,Bonzom:2012bg,Baratin:2014be} renormalizability properties \cite{BenGeloun:2013fw,Carrozza:2014bh} and phase structure \cite{Baratin:2014be,Benedetti:2014uc}.


\bibliographystyle{JHEP}
\bibliography{polyhedralGFT}

\end{document}

%% file: macros.tex
\newcommand{\fig}[1]{fig.\,\ref{fig:#1}}

\usetikzlibrary{arrows, positioning, patterns, shapes, external}
\definecolor{jgreen}{rgb}{0,0.7,0}

\tikzstyle{c} =	[coordinate]
\tikzstyle{v} = 	[circle, draw=black, line width=.5pt, fill=black, inner sep=0pt, minimum size=1.5mm]
\tikzstyle{vb} =	[circle, draw=black, line width=.5pt, fill=red, inner sep=0pt, minimum size=1.5mm]
\tikzstyle{vh} =	[circle, draw=black, line width=.5pt, fill=blue, inner sep=0pt, minimum size=1.5mm]
\tikzstyle{vs} =	[circle, draw=black, line width=.5pt, fill=jgreen, inner sep=0pt, minimum size=1.5mm]
\tikzstyle{e} =	[draw=red,line width=1.6pt]
\tikzstyle{eb} =	[draw=jgreen,line width=1.6pt]
\tikzstyle{eh} =	[dashed]
\tikzstyle{es} =	[draw=blue,line width=1.6pt]
\tikzstyle{f} = 	[line width=0.01pt,dotted,fill=blue, fill opacity=.1]
\tikzstyle{cs} = 	[draw=none,fill=red, fill opacity=.7]
\tikzstyle{bb} =	[dashed,line width=1.4pt]
\tikzstyle{bh} =	[dotted,line width=1pt]
\tikzstyle{h} = 	[ellipse, inner sep=0.1pt, draw=black]
\tikzstyle{hv} = 	[ellipse, inner sep=0.1pt, draw=red]
\tikzstyle{he} = 	[ellipse, inner sep=0.1pt, draw=blue]
\tikzstyle{hb} = 	[ellipse, inner sep=0.1pt, draw=jgreen]

\tikzstyle{venn1} = [fill=gray!20!white, draw=black, pattern=north east lines]
\tikzstyle{venn2} = [fill=gray!20!white, draw=black, pattern=north west lines]

\newcommand{\gft}{\textsc{gft}}
\newcommand{\sub}{\textsc{mf}} 

\newcommand{\std}{D}				

\newcommand{\bs}{\partial}			

\newcommand{\Vb}{\overline{\mathcal{V}}}
\newcommand{\vb}{{\bar v}}

\newcommand{\vh}{{\hat v}}

\newcommand{\gm}{\gamma}			

\newcommand{\copies}{k}

\newcommand{\loopless}{\textsc{l}}
\newcommand{\rl}{{\copies,\loopless}}

\newcommand{\simplicial}{\textsc{s}}
\newcommand{\rs}{{\copies,\simplicial}}
\newcommand{\drs}{{{\std},\simplicial}}

\newcommand{\lnb}{\textsc{l}\textrm{-}\textsc{nb}}
\newcommand{\snb}{\textsc{s}\textrm{-}\textsc{nb}}

\newcommand{\qbra}{\langle}
\newcommand{\qket}{\rangle}

\newcommand{\fdiagram}{\Gamma}
\newcommand{\sym}{\textrm{sym}}

\newcommand{\bbg}{\mathfrak{b}}
\newcommand{\bbgs}{\mathfrak{B}}

\newcommand{\bbgst}{\widetilde{\mathfrak{B}}}

\newcommand{\sfa}{\mathfrak{a}}

\newcommand{\bp}{{\mathfrak{p}}}
\newcommand{\bps}{\mathfrak{P}}

\newcommand{\bpst}{\widetilde{\mathfrak{P}}}

\newcommand{\sfr}{\mathfrak{m}}
\newcommand{\sfrs}{\mathfrak{M}}
\newcommand{\sfrt}{\widetilde{\mathfrak{m}}}
\newcommand{\sfrst}{\widetilde{\mathfrak{M}}}

\newcommand{\Vbar}{\overline{\mathcal{V}}}

\newcommand{\Dcal}{\mathcal{D}}

\newcommand{\Kbb}{\mathbf{K}}
\newcommand{\Pbb}{\mathbf{P}}

\newcommand{\Vbb}{\mathbf{V}}

\newcommand{\Kbbb}{\overline{\mathbf{K}}}

\renewcommand\[{\begin{equation}}
\renewcommand\]{\end{equation}}
\newcommand{\ba}{\begin{eqnarray}}
\newcommand{\ea}{\end{eqnarray}}

\newcommand{\N}{\mathbb N}

\newcommand{\R}{\mathbb R}

\newcommand{\tr}{\mathrm{tr}}
\def\d{\mathrm{d}}

\mathchardef\ordinarycolon\mathcode`\:
\mathcode`\:=\string"8000
\begingroup \catcode`\:=\active
  \gdef:{\mathrel{\mathop\ordinarycolon}}
\endgroup

\newcommand{\ra}{\rightarrow}

\newcommand{\In}{\subset}

\def\ie{{i.e.}~}

\newcommand{\ORT}{Oriti:2015kv}

\newcommand{\lqgT}{Thiemann:2007wt}

\newcommand{\KKL}{Kaminski:2010ba}

\newcommand{\EPRL}{
Engle:2008fj}
\newcommand{\FK}{Freidel:2008fv}
\newcommand{\BO}{Baratin:2012br}